# Multi-Frequency *GPR* Microwave Imaging of Sparse Targets Through a Multi-Task Bayesian Compressive Sensing Approach

Marco Salucci[1] and Nicola Anselmi[1]
[1]CNIT - "University of Trento" ELEDIA Research Unit
Via Sommarive 9, I-38123 Trento, Italy
E-mail: {marco.salucci, nicola.anselmi.1}@unitn.it

*Abstract* - An innovative inverse scattering (*IS*) method is proposed for the quantitative imaging of pixel-sparse scatterers buried within a lossy half-space. On the one hand, such an approach leverages on the wide-band nature of ground penetrating radar (*GPR*) data by jointly processing the multi-frequency (*MF*) spectral components of the collected radargrams. On the other hand, it enforces *sparsity priors* on the problem unknowns to yield regularized solutions of the fully non-linear scattering equations. Towards this end, a multi-task Bayesian Compressive Sensing (*MT-BCS*) methodology is adopted and suitably customized to take full advantage of the available frequency diversity and of the *a-priori* information on the class of imaged targets. Representative results are reported to assess the proposed *MF-MT-BCS* strategy also in comparison with competitive state-of-the-art alternatives.

*Introduction*: During the last decades, many efforts have been devoted to the development of microwave imaging (*MI*) techniques for retrieving reliable and easy-to-interpret images of subsurface regions starting from the radargrams collected above the interface with a ground penetrating radar (*GPR*) [1]-[9]. The solution of the arising subsurface inverse scattering (*IS*) problem poses several challenges mainly related to the intrinsic non-linearity (*NL*) and the ill-posedness (*IP*) [10]. On the one hand, the *NL* can be avoided by introducing Born-like approximations of the scattering equations [5], provided that weak scatterers are at hand and assuming that qualitative guesses (i.e., location and shape) are sufficient for the targeted application. Otherwise, multi-resolution strategies, integrated with both deterministic [7][9] and stochastic [8] optimization techniques, proved to be effective in mitigating the *NL* by reducing the ratio between unknowns and non-redundant informative data. On the other hand, the *IP* issue can be tackled by collecting the maximum amount of information from the scattering experiments. For instance, the wide-band nature of *GPR* measurements above the interface [1] provides an intrinsic frequency diversity in the collectable data. Such an information on the scenario under test can be profitably exploited with both frequency-hopping (*FH*) [6][7] and multi-frequency (*MF*) [8][9] *MI* techniques by processing each spectral component in a cascaded fashion or jointly, respectively. Another effective recipe against the *IP* is the use of the *a-priori* information on the class of imaged targets. As a matter of fact, Compressive Sensing (*CS*)-based techniques [6][11]-[14] faithfully retrieved sparse objects (i.e., objects representable with few non-null expansion coefficients with respect to a suitably-chosen representation basis). In such a framework, Bayesian *CS* (*BCS*) solvers have emerged as effective, computationally-fast, and also feasible tools since they do not require the compliancy of the scattering operator with the restricted isometry property (*RIP*), whose check is often computationally unaffordable [11]. Following this line of reasoning, this letter presents a novel *MF* approach for reliably, robustly, and efficiently solving the *GPR-MI* of pixel-sparse subsurface objects. The proposed approach is based on a fully non-linear contrast source (*CSI*) formulation of the scattering equations, then solved by means of a customized multi-task *BCS* (*MT-BCS*) solver [13][14] based on a joint marginal likelihood maximization strategy that enforces the correlation between multi-static/multi-view wide-band *GPR* data.

*Mathematical Formulation*: Let us consider a 2*D* half-space scenario where the investigation domain *D* is a subsurface region within a lossy soil with relative permittivity $\varepsilon_{rs}$ and conductivity $\sigma_s$ (Fig. 1). By considering a multi-static/multi-view measurement system, *D* is illuminated by *V* *z*-oriented line sources placed in an observation domain $\Omega$ at distance *H* above the interface (Fig. 1). The *v*-th ($v=1,...,V$) total electric field measured in time-domain by the *m*-th [$m=1,...,M$; $M=(V-1)$] receiver in $\Omega$, $\underline{r}_m^v$ [$\underline{r}_m^v = (x_m^v, y_m^v = H)$], at the time-instant *t* ($0 \le t \le T$), is given by [1]

$$e^v(\underline{r}_m^v, t) = e_i^v(\underline{r}_m^v, t) + e_s^v(\underline{r}_m^v, t) \quad (1)$$

where $e_i^v$ and $e_s^v$ are the incident and scattered fields, respectively, while *T* is the duration of the *GPR* probing window. Being $\Delta f$ [$\Delta f = (f_{\max} - f_{\min})$] the 3 [dB] bandwidth of the transmitted waveform, the scattered field at the *p*-th ($p=1,...,P$) frequency, $f_p$ [$f_p = f_{\min} + (p-1)\Delta f / (P-1)$] turns out to be

$$E_{s,p}^v(\underline{r}_m^v) = \Phi_p\{e^v(\underline{r}_m^v, t) - e_i^v(\underline{r}_m^v, t)\}; \quad m=1,...,M; \quad v=1,...,V \quad (2)$$

where $\Phi_p\{a(\underline{r},t)\} = \int_{-\infty}^{+\infty} a(\underline{r},t) \exp(-j2\pi f_p) dt$ is the Fourier transform. Moreover, it is related to the contrast function, $\tau_p(\underline{r})$

$$\tau_p(\underline{r}) = [\varepsilon_r(\underline{r}) - \varepsilon_{rs}] + j\left[\frac{\sigma_s - \sigma(\underline{r})}{2\pi f_p \varepsilon_0}\right], \quad (3)$$

modeling the unknown dielectric distribution in the investigation domain *D* at $f_p$ ($p=1,...,P$), by the data equation [1]

$$E_{s,p}^v(\underline{r}_m^v) = \int_D G_p^v(\underline{r}_m^v, \underline{r}) J_p^v(\underline{r}) d\underline{r} \quad (4)$$

where $G_p^v(\underline{r}_m^v, \underline{r})$ is the half-space Green's function [1], while $J_p^v(\underline{r})$

[$J_p^v(\underline{r}) = \tau_p(\underline{r}) E_p^v(\underline{r}) = \tau_p(\underline{r}) \Phi_p\{e^v(\underline{r},t)\}$] is the $v$-th ($v=1,...,V$) equivalent current induced within the investigation domain.

*Inverse Problem Solution Approach*: To numerically solve the inverse problem at hand, the equation (4) is first recast into the following matrix expression

$$\underline{\xi}_p^v = \underline{\underline{\Psi}}_p^v \underline{\upsilon}_p^v \qquad (5)$$

by partitioning $D$ into $N$ square sub-domains centered at $\{\underline{r}_n; n=1,...,N\}$ so that $\underline{\xi}_p^v = [\Re(\underline{E}_{s,p}^v), \Im(\underline{E}_{s,p}^v)]^T$ being $\underline{E}_{s,p}^v = \{E_{s,p}^v(\underline{r}_m^v); m=1,...,M\}$, $\underline{\upsilon}_p^v = [\Re(\underline{J}_p^v), \Im(\underline{J}_p^v)]^T$ being $\underline{J}_p^v = \{J_p^v(\underline{r}_n); n=1,...,N\}$, and

$$\underline{\underline{\Psi}}_p^v = \begin{bmatrix} \Re(\underline{\underline{G}}_p^v) & -\Im(\underline{\underline{G}}_p^v) \\ \Im(\underline{\underline{G}}_p^v) & \Re(\underline{\underline{G}}_p^v) \end{bmatrix}, \qquad (6)$$

$\underline{\underline{G}}_p^v$ being the $(v,p)$-th ($M \times N$) half-space Green's matrix [1], while $.^T$ stands for the transpose operator and $\Re(.)/\Im(.)$ denotes the real/imaginary part. Successively the solution of (5) is found with a customized multi-frequency multi-task *BCS* (*MF-MT-BCS*) technique [14] by jointly enforcing the spatial sparsity of the unknown components of the equivalent currents, $\{\underline{\upsilon}_p^v; v=1,...,V; p=1,...,P\}$ [$\underline{\tilde{\upsilon}}_p^v = \{\tilde{\upsilon}_{p,n}^v; n=1,...,2N\}$; $\tilde{\upsilon}_{p,n}^v = \Re\{\tilde{J}_p^v(\underline{r}_n)\}$ and $\tilde{\upsilon}_{p,(n+N)}^v = \Im\{\tilde{J}_p^v(\underline{r}_n)\}$], and their correlation among the different illuminations and spectral components, the number of "tasks" solved in parallel being equal to $L = (V \times P)$. More specifically, the $v$-th ($v=1,...,V$) equivalent current at the $p$-th ($p=1,...,P$) frequency is computed as

$$\underline{\tilde{\upsilon}}_p^v = \left[ diag(\underline{\tilde{\alpha}}) + (\underline{\underline{\Psi}}_p^v)^H \underline{\underline{\Psi}}_p^v \right]^{-1} (\underline{\underline{\Psi}}_p^v)^H \underline{\xi}_p^v \qquad (7)$$

by applying a fast relevant vector machine (*RVM*) method [14] to solve the following optimization problem

$$\underline{\tilde{\alpha}} = \arg\left\{ \max_{\underline{\alpha}} \left[ -0.5 \sum_{p=1}^{P} \sum_{v=1}^{V} (2M + 2\delta_1) \log\left[ (\underline{\xi}_p^v)^H (\underline{U}_p^v)^{-1} \underline{\xi}_p^v + 2\delta_2 \right] + \log|\underline{U}_p^v| \right] \right\} \qquad (8)$$

for retrieving the set of $2N$ hyper-parameters $\underline{\alpha} = \{\alpha_n; n=1,...,2N\}$ shared among the $V$ views and $P$ frequencies. In (8), $\underline{\underline{U}}_p^v = \underline{\underline{I}} + \underline{\underline{\Psi}}_p^v [diag(\underline{\alpha})]^{-1} (\underline{\underline{\Psi}}_p^v)^H$, $\underline{\underline{I}}$ being the identity matrix, while $\delta_1$ and $\delta_2$ are *BCS* control parameters. Moreover, $.^H$ and $|.|$ indicates the conjugate transpose and the determinant, respectively. Finally, the contrast distribution ($n=1,...,N$) at the central frequency, $f_c$ [$f_c = (f_{min} + f_{max})/2$], is derived as

$$\tilde{\tau}(\underline{r}_n) = \frac{1}{P} \sum_{p=1}^{P} \left( \Re\{\tilde{\tau}_p(\underline{r}_n)\} + j \frac{f_p}{f_c} \Im\{\tilde{\tau}_p(\underline{r}_n)\} \right) \qquad (9)$$

where

$$\tilde{\tau}_p(\underline{r}_n) = \frac{1}{V}\left(\sum_{v=1}^{V} \frac{\tilde{J}_p^v(\underline{r}_n)}{\tilde{E}_p^v(\underline{r}_n)}\right), \qquad (10)$$

$\tilde{J}_p^v(\underline{r}_n) = \tilde{v}_{p,n}^v + j\tilde{v}_{p,(n+N)}^v$ and $\tilde{E}_p^v(\underline{r}_n)$ being the $(v,p)$-th ($v=1,...,V$; $p=1,...,P$) retrieved current and the corresponding total electric field in the $n$-th ($n=1,...,N$) cell of the investigation domain ($\underline{r}_n \in D$), respectively.

*Numerical Assessment:* To assess the proposed *MF-MT-BCS* approach, representative numerical results are shown and discussed in this Section. A square investigation domain *D* of side 0.8 [m] buried in a medium with $\varepsilon_{rs} = 4.0$ and $\sigma_s = 10^{-3}$ [S/m] [9] has been considered as reference benchmark scenario. Moreover, a set of $V = 20$ sources and $M = 19$ probes, located in an observation domain $\Omega$ placed at $H = 0.1$ [m] above the interface (Fig. 1), has been chosen for the sensing setup to collect the time-domain *GPR* radargrams. These latter have been simulated with the *GPRMax2D* SW [15], while the scattered spectrum has been sampled at $P = 9$ uniformly-spaced frequencies within the 3 [dB] band $(f_{min}, f_{max}) = (200, 600)$ [MHz] ($\Delta f = 400$ [MHz], $f_c = 400$ [MHz]) [6]. As for the setting of the *MF-MT-BCS* control parameters (8), the optimal trade-off values $(\delta_1, \delta_2) = (6 \times 10^{-1}, 9 \times 10^{-5})$ have been derived from a preliminary calibration performed by blurring the time-domain total field data samples with different levels of white Gaussian noise.

The first test case is concerned with the "*Two-Bars*" scattering profile of Fig. 2(*a*) ($\varepsilon_r = 5, \sigma = 10^{-3}$ [S/m] $\Rightarrow \tau = 1.0$). The *MF-MT-BCS* data inversion gives a very accurate image of *D* independently on the data signal-to-noise ratio (*SNR*) and it faithfully recovers the support as well as the contrast value of the two buried scatterers [Figs. 2(*b*)-2(*c*) vs. Fig. 2(*a*)]. To better point out the advantage of jointly processing all spectral components of the scattered field, as done by the proposed *MF* inversion scheme, the results of two *FH*-based state-of-art *BCS* solution strategies are reported in Fig. 2 for comparison purposes. It is worth reminding that these methods process each *p*-th ($p=1,...,P$) frequency in a cascaded fashion, from the lowest to the highest one, by either enforcing the correlation between multiple views ($L = V$ - *FH-MT-BCS* method [6]) or considering each view as a single task ($L = 1$ - *FH-ST-BCS* method [6]). As it can be observed, the *MF-MT-BCS* outperforms both *FH* strategies, the worst inversion being performed by the *FH-ST-BCS* [Figs. 2(*f*)-2(*g*)]. Such outcomes are quantitatively confirmed by the values of the total error, $\Xi_{tot}$, computed as in [9] and reported in Fig. 3 versus the *SNR*. The *MF-MT-BCS* does not only provide the lowest errors, but it is also significantly more robust against the data noise since, for instance, $\Xi_{tot}\big|_{MF-MT-BCS}^{SNR=35dB} / \Xi_{tot}\big|_{FH-MT-BCS}^{SNR=35dB} = 2.5 \times 10^{-1}$ [Fig. 2(*e*) vs. Fig. 2(*c*)] and $\Xi_{tot}\big|_{MF-MT-BCS}^{SNR=35dB} / \Xi_{tot}\big|_{FH-ST-BCS}^{SNR=35dB} = 3.0 \times 10^{-2}$ [Fig. 2(*g*) vs. Fig. 2(*c*)] in the most critical working conditions (i.e., $SNR = 35$ [dB]).

Similar conclusions can be drawn also when dealing with a more complex-shaped scatterer. As a matter of fact, the "*S-shaped*" object [$\tau = 1.0$, Fig. 4(*a*)] has been imaged by the *MF-MT-BCS* [Fig. 4(*b*) vs. Fig. 4(*a*)] remarkably better than the *FH-MT-BCS* [Fig. 4(*c*)] and the *FH-ST-BCS* [Fig. 4(*d*)], both *FH* methods failing in retrieving the actual support of the scatterer.

The *MF-MT-BCS* is more effective to recover objects with a higher conductivity than the hosting medium, as well. Indeed, despite the increased complexity due to the presence of a non-null imaginary part of the contrast and the non-negligible amount of noise, it is the only method able to provide an accurate guess of both the real part [Fig. 5(*c*) vs. Fig. 5(*a*)] and the imaginary one [Fig. 5(*d*) vs. Fig. 5(*b*)] of the "*Diagonal*" scatterer [$\varepsilon_r = 5$, $\sigma = 10^{-2}$ [S/m] $\Rightarrow \tau = 1.0 - j0.4$ - Figs. 5(*a*)-5(*b*)]. Besides the pictorial representations in Fig. 5, the performance of each inversion method have been quantified in terms of the total, the internal (i.e., within the target support), and the external (i.e., in the background) errors [9], the corresponding values being reported in Tab. I.

Finally, it is worth pointing out the higher efficiency exhibited by the *MF-MT-BCS* thanks to the "one-shot" inversion of all *P* frequency components of the *GPR* spectrum. As a representative example, let us consider that the reduction of the inversion time on a standard laptop with Intel(R) Core(TM) i5-8250U CPU @ 1.60GHz and 16 [GB] of RAM amounts to $\Delta t|_{FH-MT-BCS} / \Delta t|_{MF-MT-BCS} = 22.1$ and $\Delta t|_{FH-ST-BCS} / \Delta t|_{MF-MT-BCS} = 85.8$, respectively (Tab. I).

*Conclusion:* A novel sparsity-promoting strategy has been proposed to effectively solve the *2D GPR-MI* problem. Thanks to the adopted *MF* strategy, the *MF-MT-BCS* method allows a computationally-efficient exploitation of the frequency-diversity of the *GPR* data by correlating all the multi-chromatic components extracted from the measured radargrams. As a result, it outperforms available *FH*-based solution strategies formulated within the *BCS* framework by exhibiting remarkably higher accuracy, robustness, and computational efficiency.

*Acknowledgments*: This work has been partially supported by the Italian Ministry of Education, University, and Research within the Program PRIN 2017 (CUP: E64I19002530001) for the Project CYBERPHYSICAL ELECTROMAGNETIC VISION: Context-Aware Electromagnetic Sensing and Smart Reaction (EMvisioning) (Grant no. 2017HZJXSZ), within the Program "Smart cities and communities and Social Innovation" for the Project "SMARTOUR - Piattaforma Intelligente per il Turismo" (Grant no. SCN_00166), and within the Program "Smart cities and communities and Social Innovation" (CUP: E44G14000060008) for the Project "WATERTECH" (Grant no. SCN_00489).

**References**

[1]   Persico R. Introduction to ground penetrating radar: inverse scattering and data processing. London: Wiley-IEEE Press; 2014.
[2]   Baussard A, Miller EL, Lesselier D. Adaptive multiscale reconstruction of buried objects. Inverse Prob. 2004;20:S1-S15.

[3]    Li M, Abubakar A, Habashy TM. Application of a two-and-half dimensional model-based algorithm to crosswell electromagnetic data inversion. Inverse Prob. 2010;26(7):1–17.

[4]    Abubakar A, Habashy TM, Li M, Liu J. Inversion algorithm for large-scale geophysical electromagnetic measurements. Inverse Prob. 2009;25(12):1–30.

[5]    Meincke P. Linear GPR inversion for lossy soil and a planar air-soil interface. IEEE Trans. Geosci. Remote Sens. 2001;39(12):2713-2721.

[6]    Salucci M, Gelmini A, Poli L, Oliveri G, Massa A. Progressive compressive sensing for exploiting frequency-diversity in GPR imaging. J. Electromagn. Waves Appl. 2018;32(9): 1164-1193.

[7]    Salucci M, Oliveri G, Massa A. GPR prospecting through an inverse scattering frequency-hopping multi-focusing approach. IEEE Trans Geosci Remote Sens. 2015;53(12):6573–6592.

[8]    Salucci M, Poli L, Anselmi N, Massa A. Multifrequency particle swarm optimization for enhanced multiresolution GPR microwave imaging. IEEE Trans Geosci Remote Sens. 2017;55(3):1305–1317.

[9]    Salucci M, Estatico C, Fedeli A, Oliveri G, Pastorino M, Povoli S, Randazzo A, Rocca P. 2D TM GPR imaging through a multi-scaling multi-frequency approach in Lp-spaces. IEEE Trans Geosci Remote Sens. 2021 (doi: 10.1109/TGRS.2021.3049423).

[10]   Chen X. Computational methods for electromagnetic inverse scattering. Wiley; 2018.

[11]   Massa A, Rocca P, Oliveri G. Compressive sensing in electromagnetics - A review. IEEE Antennas Propag Mag. 2015;57(1):224–238.

[12]   Ambrosanio M, Pascazio V. A compressive-sensing-based approach for the detection and characterization of buried objects. IEEE J Selected Topics Appl Earth Observ Remote Sens. 2015;8(7):3386–3395.

[13]   Sun Y, Qu L, Zhang S, Yin Y. MT-BCS-based two-dimensional diffraction tomographic GPR imaging algorithm with multiview-multistatic configuration. IEEE Geosci Remote Sens Lett. 2016;13(6):831–835.

[14]   Ji S, Dunson D, Carin L. Multitask compressive sensing. IEEE Trans. Signal Process. 2009;57(1):92–106.

[15]   Warren C, Giannopoulos A, Giannakis I. gprMax: Open source software to simulate electromagnetic wave propagation for Ground Penetrating Radar. Computer Physics Communications. 2016;209:163-170.

**Figure Captions**

**Figure 1.**  Geometry of the *2D-TM GPR-MI* problem

**Figure 2.**  *Numerical Assessment ("Two-Bars" Scatterer, $\tau = 1.0$, $N = 400$ ) - Actual (a) and retrieved (b)-(g) dielectric profile by (b)(c) the MF-MT-BCS, (d)(e) the FH-MT-BCS, and (f)(g) the FH-ST-BCS when processing noisy data at (b)(d)(f) $SNR = 50$ [dB] and (c)(e)(g) $SNR = 35$ [dB].*

**Figure 3.**  *Numerical Assessment ("Two-Bars" Scatterer, $\tau = 1.0$, $N = 400$, $SNR \in [35, 55]$ [dB]) - Behavior of the total integral error as a function of the SNR on time-domain total field for the MF-MT-BCS, FH-MT-BCS, and FH-ST-BCS methods.*

**Figure 4.**  *Numerical Assessment ("S-Shaped" Scatterer, $\tau = 1.0$, $N = 400$, $SNR = 35$ [dB]) - Actual (a) and retrieved (b)-(d) dielectric profile by (b) the MF-MT-BCS, (c) the FH-MT-BCS, and (d) the FH-ST-BCS.*

**Figure 5.**  *Numerical Assessment ("Diagonal" Scatterer, $\tau = 1.0 - j0.4$, $N = 400$, $SNR = 35$ [dB]) - Actual (a)(b) and retrieved (c)-(h) real part (a)(c)(e)(g) and imaginary part (b)(d)(f)(h) of the contrast outputted by (c)(d) the MF-MT-BCS, (e)(f) the FH-MT-BCS, and (g)(h) the FH-ST-BCS.*

**Table Captions**

**Table I.**  *Numerical Assessment ("Diagonal" Scatterer, $\tau = 1.0 - j0.4$, $N = 400$, $SNR = 35$ [dB]) - Total, internal, and external reconstruction errors [9] and inversion time for the MF-MT-BCS, the FH-MT-BCS, and the FH-ST-BCS methods.*

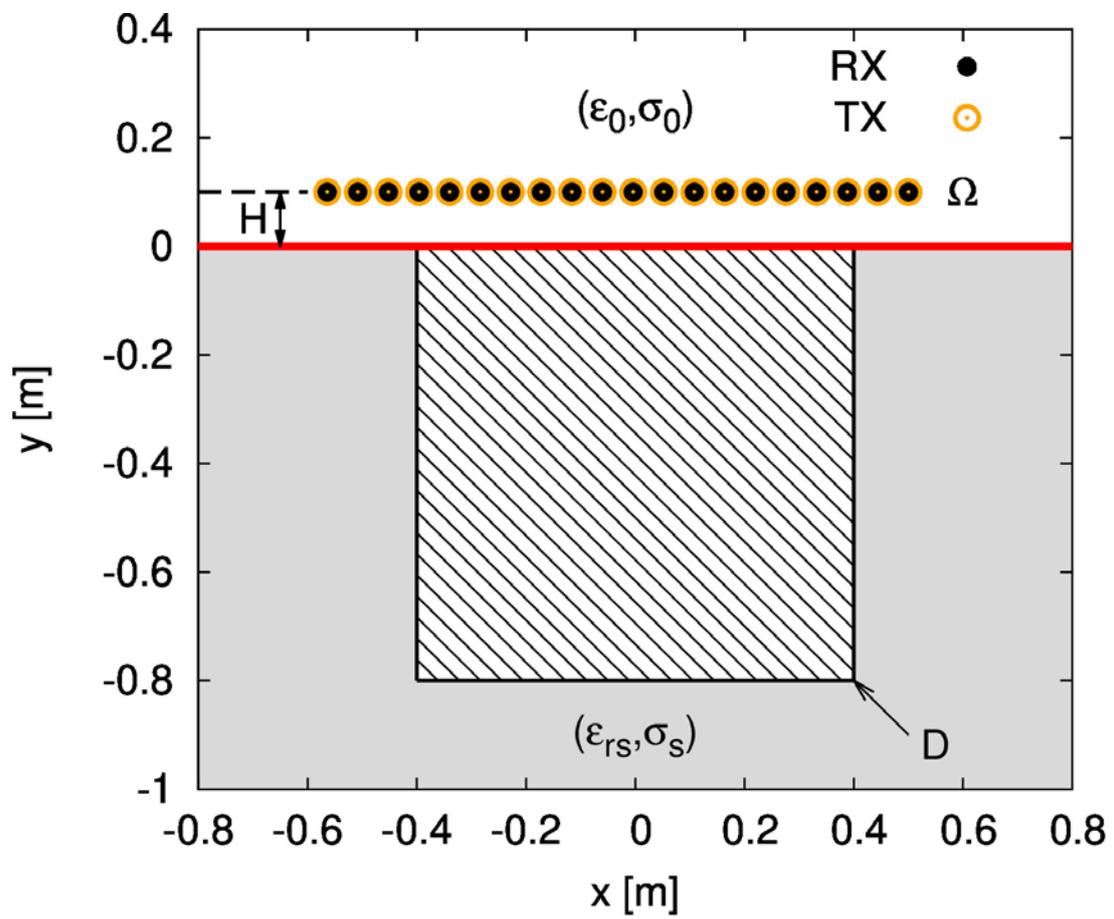

Fig. 1 – M. Salucci and N. Anselmi, "Multi-Frequency GPR Microwave Imaging of Sparse Targets …"

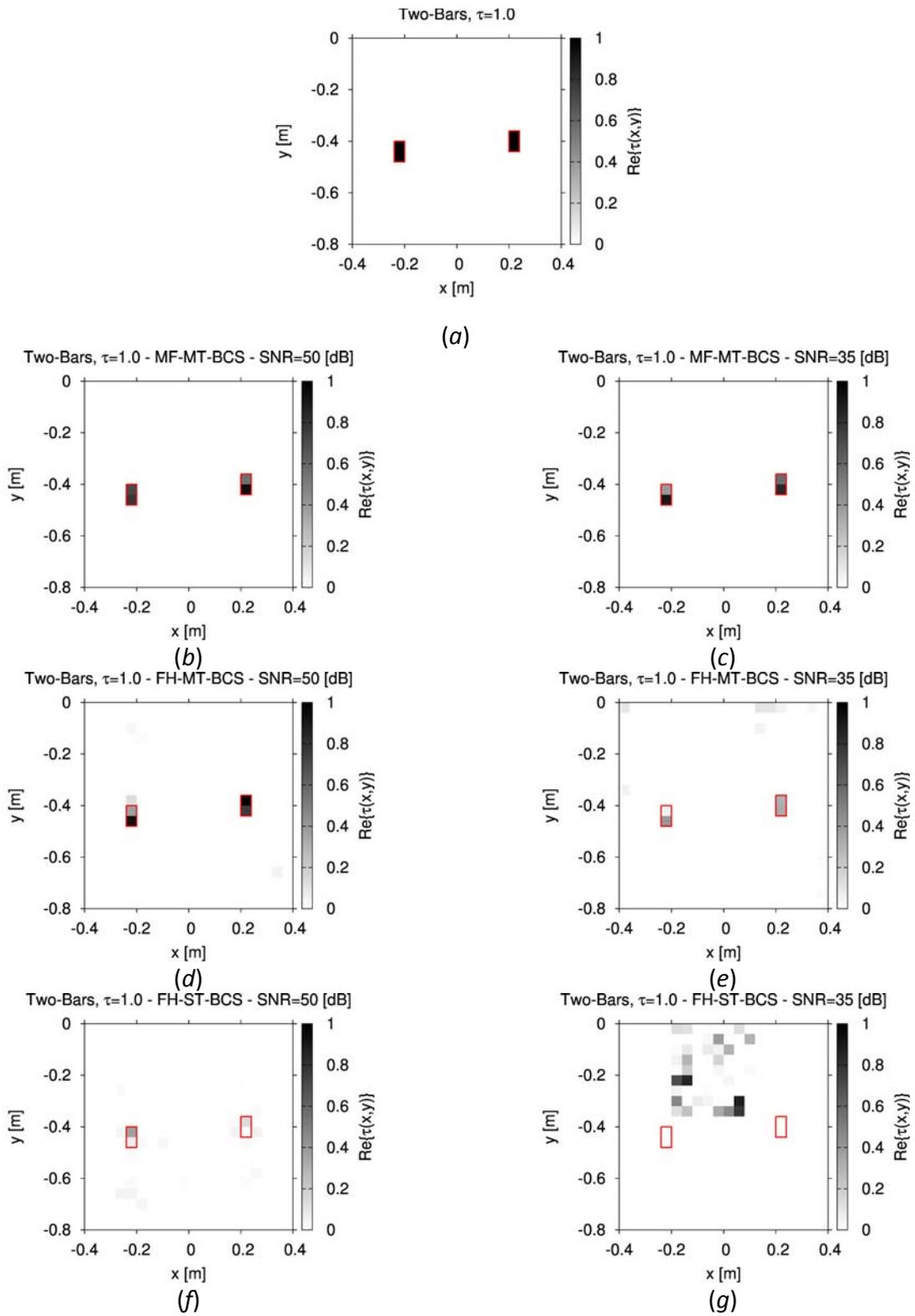

Fig. 2 – M. Salucci and N. Anselmi, "Multi-Frequency GPR Microwave Imaging of Sparse Targets …"

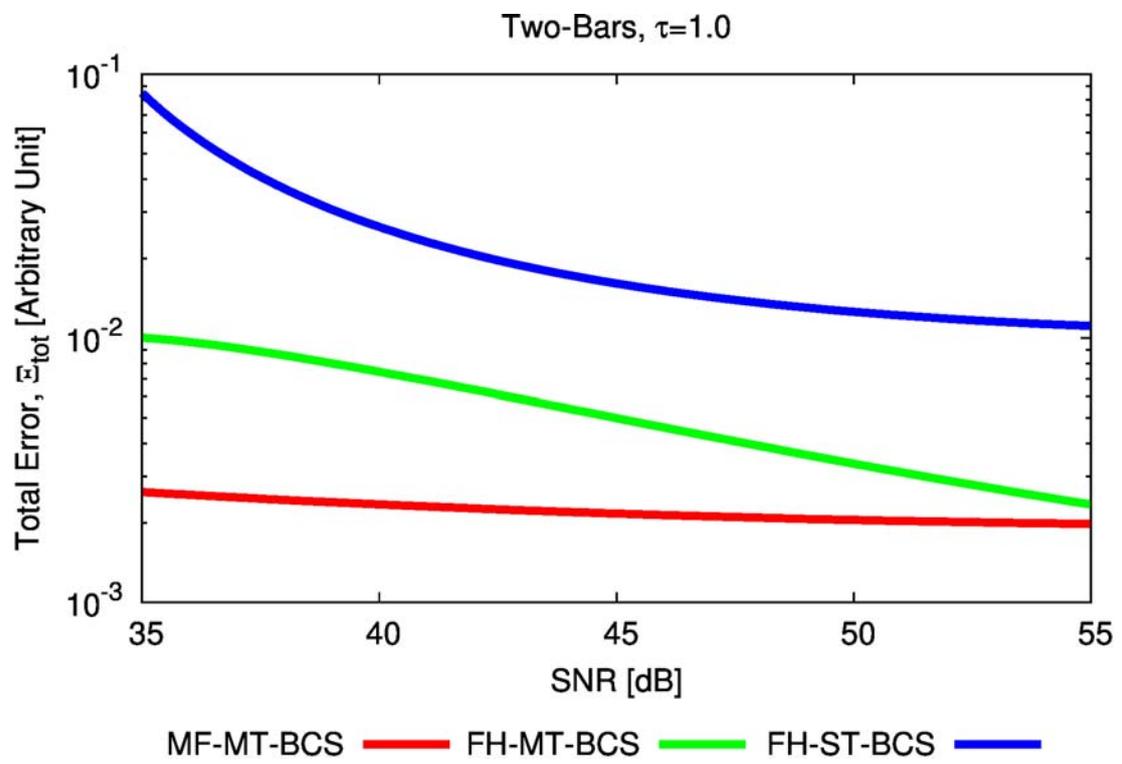

Fig. 3 – M. Salucci and N. Anselmi, "Multi-Frequency GPR Microwave Imaging of Sparse Targets …"

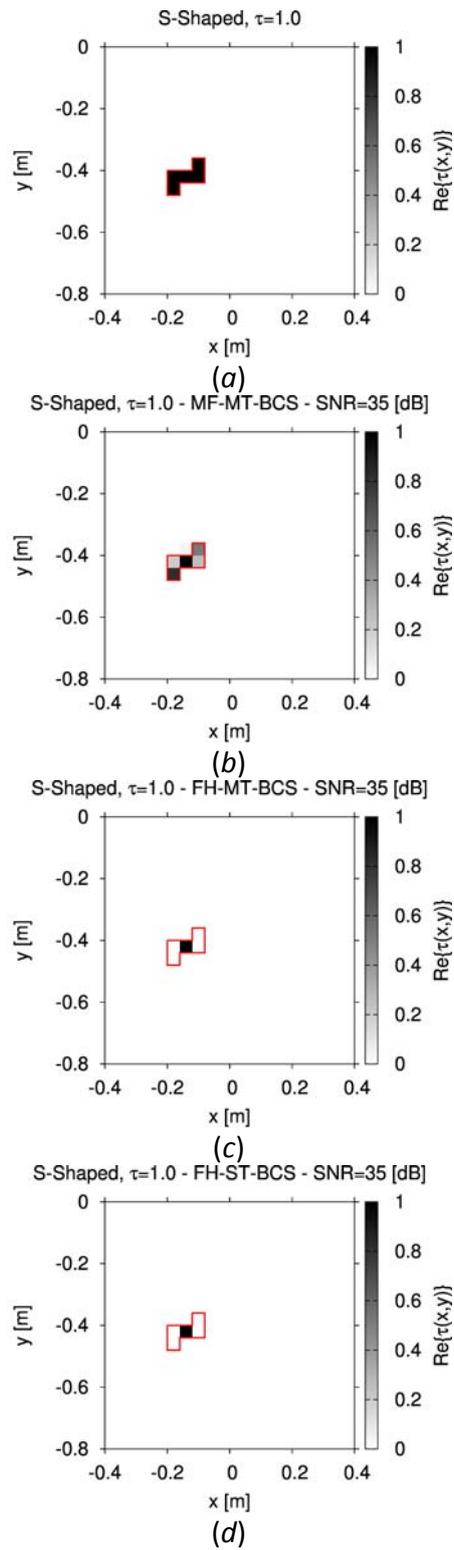

**Fig. 4** – M. Salucci and N. Anselmi, "Multi-Frequency GPR Microwave Imaging of Sparse Targets …"

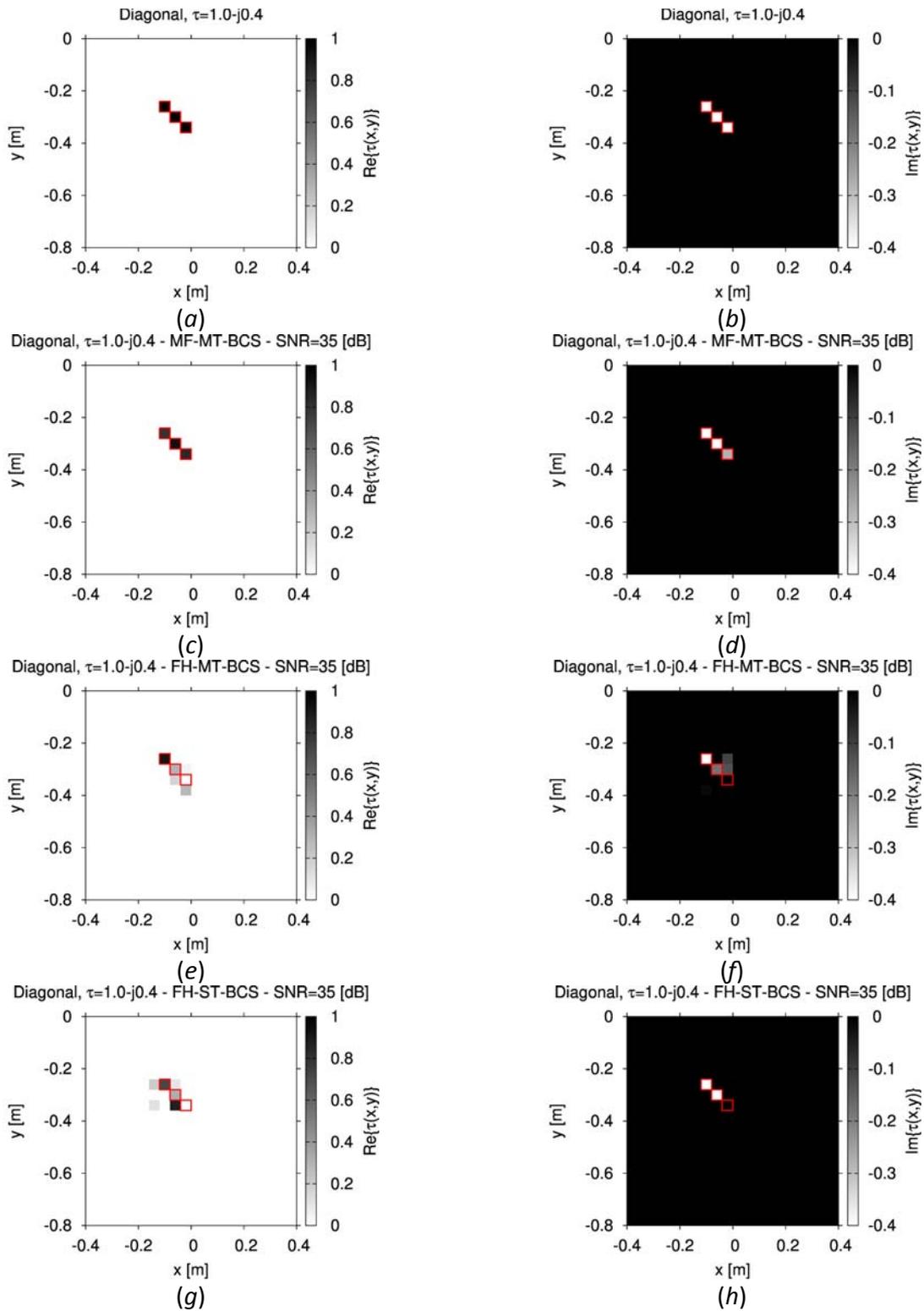

**Fig. 5** – M. Salucci and N. Anselmi, "Multi-Frequency GPR Microwave Imaging of Sparse Targets …"

|            | *MF-MT-BCS*       | *FH-MT-BCS*       | *FH-ST-BCS*       |
|------------|-------------------|-------------------|-------------------|
| $\Xi_{tot}$ | $5.94 \times 10^{-4}$ | $6.11 \times 10^{-3}$ | $1.09 \times 10^{-2}$ |
| $\Xi_{int}$ | $7.92 \times 10^{-2}$ | $3.37 \times 10^{-1}$ | $3.57 \times 10^{-1}$ |
| $\Xi_{ext}$ | 0.0               | $3.60 \times 10^{-3}$ | $8.24 \times 10^{-3}$ |
| $\Delta t$ [sec] | 3.1          | 68.5              | 266               |

**Tab. I** – M. Salucci and N. Anselmi, "Multi-Frequency GPR Microwave Imaging of Sparse Targets …"